\documentclass{article}

\usepackage{arxiv}

\usepackage[utf8]{inputenc} 
\usepackage[T1]{fontenc}    
\usepackage{hyperref}       
\usepackage{url}            
\usepackage{booktabs}       
\usepackage{amsfonts}       
\usepackage{nicefrac}       
\usepackage{microtype}      
\usepackage{url,hyperref,lineno,microtype,subcaption}
\usepackage{ulem}
\usepackage{amsmath}
\usepackage{multirow}
\usepackage{graphicx}
\usepackage{subcaption}
\usepackage{fancyvrb}
\usepackage{xcolor}
\title{Application of Seq2Seq Models on Code Correction}

\author{
  Shan Huang\\
  Department of Physics\\
  Boston University\\
  Boston, MA, 02215 \\
  \texttt{sh2015@bu.edu} \\
   \And
 Xiao Zhou \\
  Department of Computer Science\\
  Boston University\\
  Boston, MA, 02215 \\
  \texttt{zhouxiao@bu.edu} \\
  \AND
  Sang Chin \\
  Department of Computer Science \\
  Boston University \\
  Boston, MA, 02215 \\
  \texttt{spchin@bu.edu} \\
}

\begin{document}

\DefineVerbatimEnvironment%
{MyVerbatim}{Verbatim}
{frame=lines,framerule=0.8mm, fontsize=\small}

\maketitle

\begin{abstract}
We apply various seq2seq models on programming language correction tasks on Juliet Test Suite for C/C++ and Java of Software Assurance Reference Datasets(SARD), and achieve 75\%(for C/C++) and 56\%(for Java) repair rates on these tasks. We introduce Pyramid Encoder in these seq2seq models, which largely increases the computational efficiency and memory efficiency, while remain similar repair rate to their non-pyramid counterparts. We successfully carry out error type classification task on ITC benchmark examples (with only 685 code instances) using transfer learning with models pre-trained on Juliet Test Suite, pointing out a novel way of processing small programing language datasets.
\end{abstract}

\keywords{programing language correction \and  seq2seq \and pyramid encoder \and attention mechanism \and transfer learning}

\section{Introduction}
Programming language correction (PLC), which can provide suggestions for people to debug code, identify potential flaws in a program, and help programmers to improve their coding skills, has been an important topic in Natural Language Processing (NLP) area. At present, most work in PLC used rule-based methods [\cite{jet2016} \cite{syn2016} \cite{gc2016} \cite{ge2016} \cite{singh2013automated}], using static analyzers, code transformations or control flow to identify and correct common suspicious code patterns or potential bugs. Machine learning methods has been a minority in PLC, examples are \cite{gupta2018deep}, who used reinforcement learning based on compiler error messages, and \cite{pu2016sk_p}, who used LSTM on correcting students' programming exercises in MOOCs.

\begin{figure}
    \centering
    \includegraphics[width=0.5\textwidth]{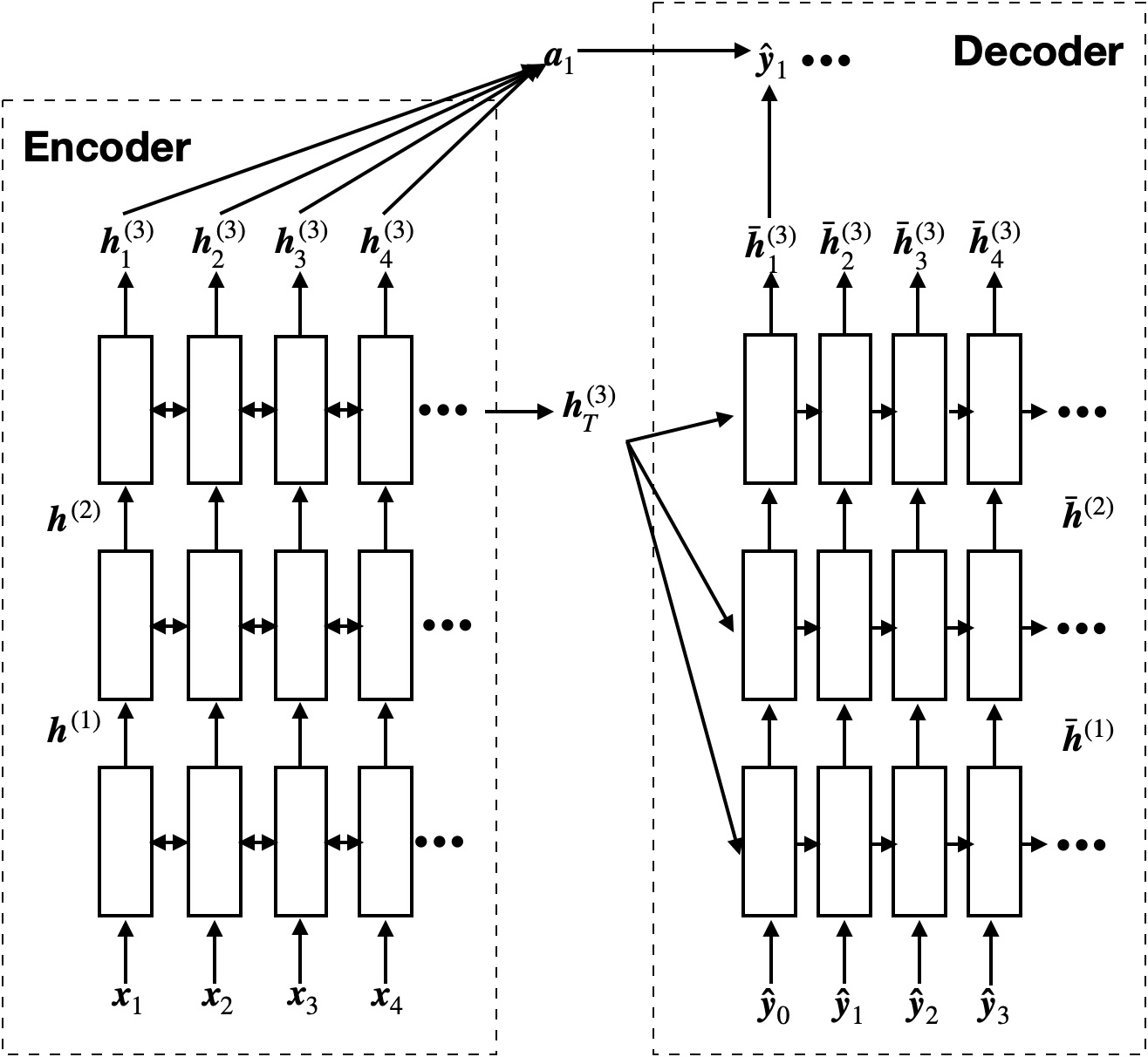}
    \caption{Model structure of a 3-layer seq2seq model with attention. The $i^{\mathrm{th}}$ layer takes the output of the previous layer ($\boldsymbol{h}^{(i-1)}$)as its input. $\boldsymbol{a}$ is the context vector, which can be calculated using different attention mechanisms.}
    \label{fig:seq2seq}
\end{figure}

Seq2seq (abbreviation of sequence to sequence) model is a group of neural network based model. It usually consist of an encoder and a decoder. The encoder takes a sequence as input, and produce an encoded representation of the input sequence. The decoder takes this representation and produce an output sequence. Therefore, this architecture is very suitable for Natural Language Processing (NLP), because language is consist of sequence of words. It is widely used in Neural Machine Translation, Text Generation, etc. An example of a seq2seq model structure is shown in Figure~\ref{fig:seq2seq}.

In natural language correction (NLC), seq2seq models are the state-of-the-art methods [\cite{xie2016neural}, \cite{vaswani2017attention}]. Considering PLC and NLC are very similar in form, we apply various seq2seq models on PLC. Unlike our predecessors, we not only aim at correcting syntax errors, but also identifying potential flaws that will not be caught during compiling, for example, memory allocation errors, redundant code, etc. One of the dataset that can serve this purpose is the Juliet Test Suite, which contains over 100 Common Weakness Enumerations(CWEs), each of them contains hundreds of examples. Our results show that seq2seq models successfully repair over 70\% of the code instances if 1 beam search candidate considered, and over 90\% if 5 beam search candidates considered.

\begin{figure}
    \centering
    \includegraphics[width=0.5\textwidth]{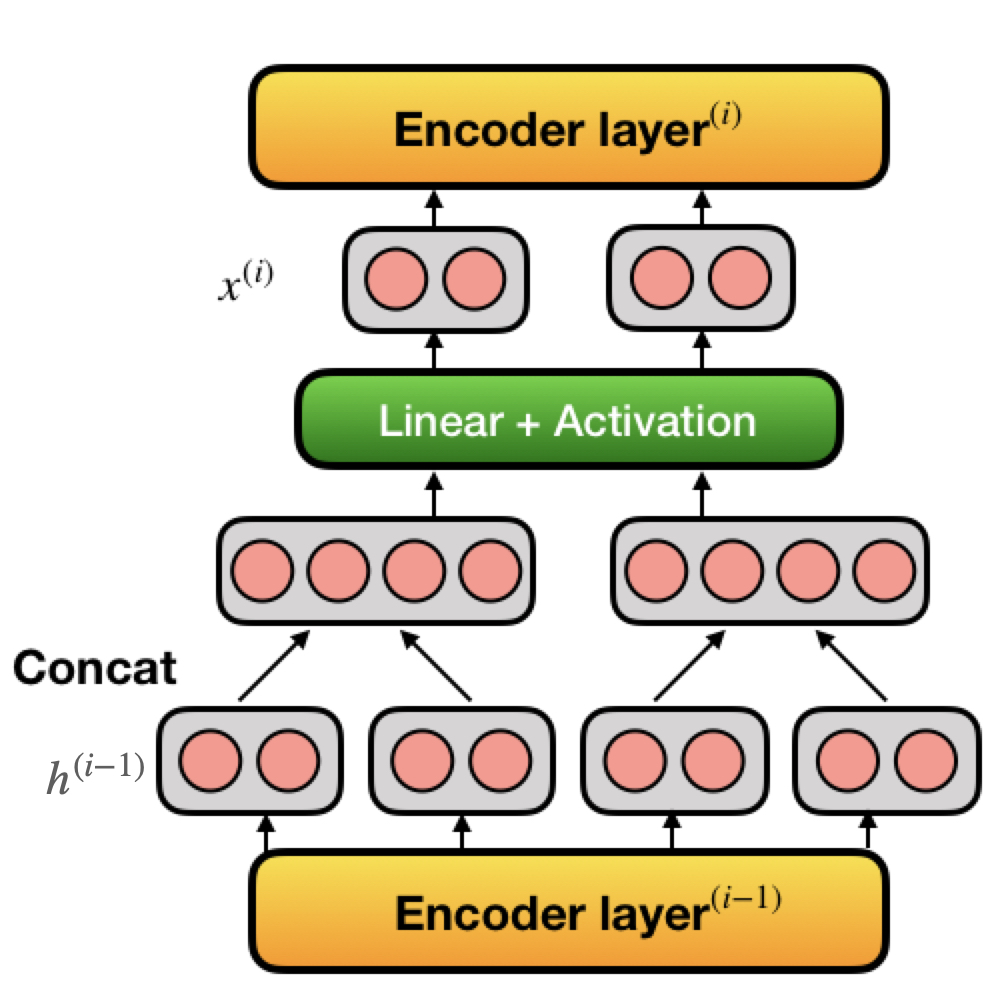}
    \caption{Visualization of Pyramid Encoder in multi-layer seq2seq models. Pyramid Encoder reduces length of input sequence by half in every encoding layer. $h^{(i-1)}$ denotes output of $(i-1)^{\text{th}}$ encoder layer and $x^{(i)}$ denotes the input of $i^{\text{th}}$ encoder layer.}
    \label{fig:Pyramid}
\end{figure}

Besides applying regular seq2seq models, we introduce pyramid encoder structure. The motivation is as follows: Unlike NLC problems, where the average length of a sentence lies around dozens of words, the average length of code instances in PLC are usually hundreds of words, which results in enormous computational cost and memory requirement. Pyramid structure aims to reduce these costs by contracting the data flow and disgard redundant information. Figure~\ref{fig:Pyramid} shows a visual representation of the pyramid encoder, it can be implemented to most of the multi-layer seq2seq learning models. In our model comparison set, Pyramid Encoder increases networks' computational efficiency by 50\% to 100\% and memory efficiency by up to 600\%, while remaining similar ability of reparation.

On the other hand, due to the privacy policies, most of the publicly available dataset are not collected from realistic program errors \& fixes, but rather are generated by artificial tools. The ones that are collected realistically are usually very small. To handle this issue, we also applied transfer learning to inherit the knowledge learned from previous dataset to boost the network's performance on smaller and noisier dataset. Details of our project are available on GitHub\footnote{See~https://github.com/b19e93n/Pyramid-Seq2Seq-Code-Correction}.

\section{Related Work}

The network models we used originated from various previous works on sequence to sequence networks. One of such networks is introduced in \cite{xie2016neural}, where the author proposed a pyramid structured bi-directional GRU to perform corrections on natural languages. They applied their model to fixing language errors on Lang-8 Corpus and CoNLL Shared Tasks, where the model achieved a BLEU score of 61.7 and 40.56 correspondingly.

Different attention mechanisms are applied in our networks. One of them is Bahdanau Attention [\cite{bahdanau2014neural}]. In this work, the content-based attention is applied with a encoder-decoder model, and achieved state-of-the-art performance on English-French translation task in 2014. The other is originally introduced by \cite{luong2015effective}, in their work, they examines two attention mechanism: one always attends to all source words and one that only looks at a subset of source words. They demonstrate that with local attention, the model is able to achieve the best BLEU score at that time. 

The multilayer structure is derived from the idea in the works by \cite{sutskever2014sequence}. The GRU cell we used is adapted from \cite{cho2014learning}. The Transformer network we use originated from \cite{vaswani2017attention}, which is a novel encoder-decoder structure that use Multi-head Attention, a positional, instead sequential, mechanism. All of the above models and methods have been proven very successful in sequence to sequence learning.

Other related works include \cite{jean2014using}, where they introduced importance sampling method (to approximate expected gradient) to use a very large target dataset without increasing the training complexity of networks. Their modified neural machine translation with attention mechanism performed well on large-vocabulary tasks (i.e. language translation), with limited training set. As in area of programming language processing, \cite{pu2016sk_p} proposed a token-based RNN model with LSTM cells, which achieved some good results in correcting student's submission in MOOCs. 

\section{Model}

\ 

\subsection{Overview}
Given a code instance, we wish to identify and correct potential flaws (if they exist) in it, which might lead to a failure in execution after successful compilation. We do not aim at correcting syntax errors, which could be handled by compilers already. 

For this purpose, we applied two major families of seq2seq models: GRU and Transformer. We use learnable embedding layers, which allows the model to recognize the relationship between different words in the vocabulary. For the encoder, we applied Pyramid Encoder, where a pyramid module is dded in between layers of regular multi-layer encoders. For the purpose of testing generality of pyramid encoder, we combined it will different attention mechanisms. 

\

\subsection{Word-level reasoning}

In language correction, character-level reasoning is a more commonly applied method~\cite{xie2016neural}. However, in code correction, we apply word-level models. The reason is that the basic building blocks of a code instance are syntax. In the field of programming language processing, out-of-vocabulary (OOV) is less a problem than in natural language due to a fixed syntax pool.

In order to prevent the model suffering from vast variation of variable names, we performed a certain degree of variable re-naming. We focused on renaming function names in our dataset while keeping other variables unchanged. This method reduced vocabulary size to $\sim$1000, and was proven to be effective in improving the performance.

We use a slightly different definition for "word" in the code correction scenario, comparing to its definition in natural language. In a code instance, any entity that is contained in the standard syntax list is considered as a "word", including a space. Details of our method of parsing the code instances to their most compact form is explained in the appendix.

\ 

\subsection{Pyramid Encoder}

Given a multi-layer seq2seq encoder, its input at $i^{\mathrm{th}}$ layer at step $t$ is $\boldsymbol{x}_t^{(i)}$, and output is $\boldsymbol{h}_t^{(i)}$:
\begin{equation}
    \boldsymbol{h}_t^{(i)} = \mathrm{Layer}^{(i)}(\boldsymbol{x}_t^{(i)})
\end{equation}
in standard seq2seq models, the output of the $i^{\mathrm{th}}$ layer $\boldsymbol{h}^{(i)}$ is directly used as input of of the $i+1^{\mathrm{th}}$ layer, $\boldsymbol{x}^{(i+1)}$:
\begin{equation}
    \boldsymbol{x}_t^{(i+1)} = \boldsymbol{h}_t^{(i)}
\end{equation}
and the time step $t = 1, 2, ..., T$, the layer number $i = 1, 2, ...,N$. Note that $\boldsymbol{x}_t^{(0)}$ is the embedded representation of the input instance. 

For pyramid encoder, we introduce a pyramid module in between $\boldsymbol{h}^{(i)}$ and $\boldsymbol{x}^{(i+1)}$ as eq. \ref{eq: pyr}:
\begin{equation}\label{eq: pyr}
    \boldsymbol{x}_{t'}^{(i+1)}= \mathrm{tanh}(\boldsymbol{W}_{\mathrm{pyr}}(\boldsymbol{h}_{2t}^{(i)}, \boldsymbol{h}_{2t + 1}^{(i)})+ b_{pyr})
\end{equation}
This module reduced the length of the input $\boldsymbol{x}^{(i)}$ by half each time it is applied. The length of final output of the encoder is $T/2^{N-1}$. One could also take a bigger window such as 3, 4, 5... depending on their needs. The hope is that Pyramid structure will extract the important information and reduce the redundant information of each of the neighboring hidden state, therefore reduce the training cost while keep the accuracy of the correction. This is conceptually similar to a convolution, but without using filters.

For our GRU models, we use multi-layer bi-directional GRU, we implemented Pyramid encoder as described first in~\cite{xie2016neural}:

\begin{align}
    \label{eq:forward}
    \boldsymbol{f}_t^{(i)} & = \mathrm{GRU} (\boldsymbol{f}_{t-1} ^{(i)}, \boldsymbol{x}_t^{(i)})\\
    \label{eq:backward}
    \boldsymbol{b}_t^{(i)} & = \mathrm{GRU} (\boldsymbol{b}_{t+1} ^{(i)}, \boldsymbol{x}_t^{(i)})\\
    \boldsymbol{h}_t^{(i)} & = \boldsymbol{f}_t^{(i)} + \boldsymbol{b}_t^{(i)}\\
    \boldsymbol{x}_{t'}^{(i + 1)} & = \mathrm{tanh}(\boldsymbol{W}_{\mathrm{pyr}}(\boldsymbol{h}^{(i)}_{2t}, \boldsymbol{h}^{(i)}_{2t + 1})+ b_{pyr})
\end{align}
where $\boldsymbol{x}_{t'}^{(i+1)}$ denotes the input to next layer, $\boldsymbol{f}_t^{(i)}$ and $\boldsymbol{b}_t^{(i)}$ denotes output from a forward and a backward GRU respectively. GRU (Gated Recurrent Unit) is a RNN (Recurrent Neural Network) type model that included gating mechanism. It can be summarized in following equations (\cite{cho2014learning}):

\begin{align}
    \boldsymbol{r}_t & = \sigma(\boldsymbol{W}_{ir}\boldsymbol{x}_t + b_{ir} + \boldsymbol{W}_{hr} \tilde{\boldsymbol{h}}_{t-1} + b_{hr}\\
    \boldsymbol{z}_t & = \sigma(\boldsymbol{W}_{iz}\boldsymbol{x}_t + b_{iz} + \boldsymbol{W}_{hz} \tilde{\boldsymbol{h}}_{t-1} + b_{hz}\\
    \boldsymbol{n}_t & = \mathrm{tanh}(\boldsymbol{W}_{in}\boldsymbol{x}_t + b_{in} + \boldsymbol{r}_t * \boldsymbol{W}_{hn} \tilde{\boldsymbol{h}}_{t-1} + b_{hn}) \\
    \tilde{\boldsymbol{h}}_t &= (1-\boldsymbol{z}_t) * \boldsymbol{n}_t + \boldsymbol{z}_t *\tilde{\boldsymbol{h}}_{t-1}
\end{align}
where $\tilde{\boldsymbol{h}}_t$ is the hidden state at step $t$, which is denoted by $\boldsymbol{f}_t$ in eq.~\ref{eq:forward} and $\boldsymbol{b}_t$ in eq.~\ref{eq:backward}. $\boldsymbol{r}_t$, $\boldsymbol{z}_t$, $\boldsymbol{n}_t$ are the reset, update and new gates, respectively. $\sigma$ is the sigmoid function.

\begin{figure}
    \centering
    \includegraphics[width=0.5\textwidth]{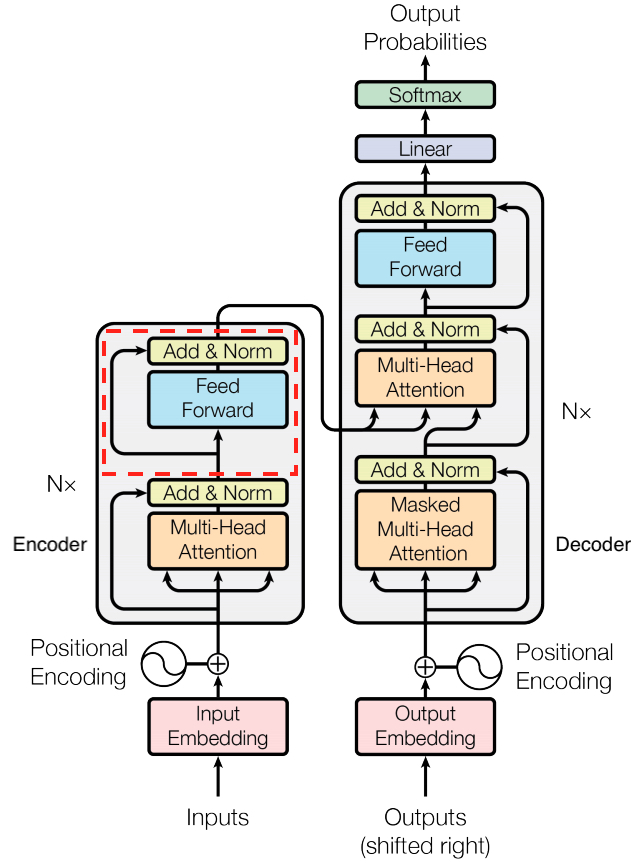}
    \caption{Structure of Transformer (\cite{vaswani2017attention}). Our pyramid encoder module replaces The Feed Forward and the Add \& Norm layer (red dashed rectangle).}
    \label{fig:TFM}
\end{figure}

Transformer is a novel family of seq2seq model that works very differently than RNN type models. In the original Transformer (see Figure~\ref{fig:TFM}), a Feed Forward layer directly takes in the output from the Multi-Head attention layer $\boldsymbol{c}_{\mathrm{att}}$, accompanied by a residual connection, shown in equation \ref{eq:TFM1}, \ref{eq:TFM2}:
\begin{align}
\label{eq:TFM1}
    \boldsymbol{c}_{\mathrm{att}}^{(i)} & = \mathrm{MultiHeadAtt}(\boldsymbol{x}^{(i)}) + \boldsymbol{x}^{(i)}\\
\label{eq:TFM2}
    \boldsymbol{x}^{(i+1)} & = \boldsymbol{c}_{\mathrm{att}}^{(i)} + \mathrm{FeedForward}(\boldsymbol{c}_{\mathrm{att}}^{(i)})
\end{align}

In our model, we concatenate the neighboring elements in $\boldsymbol{c}_{\mathrm{att}}$ before we feed it into Feed Forward layer. As a result, the dimension of the first Linear layer in Feed Forward layer has to change from $[d_{\mathrm{model}} \times d_{\mathrm{ff}}]$ to $[ 2d_{\mathrm{model}} \times d_{\mathrm{ff}}]$. Here we use the same notation as in \cite{vaswani2017attention}, where $d_{\mathrm{model}}$ is the size of input, output and attention vectors, $d_{\mathrm{ff}}$ is the number of neurons in the Feed Forward layer. The residual connection also has to be changed accordingly, we tried two different approaches, simply averaging the neighboring element (eq. \ref{eq: ave}), or concatenate the neighboring element and pass it through another affine transformation to recover its dimensions (eq. \ref{eq: aff}). For simplicity, we denote former method with subscript "ave" and the latter with subscript "aff".

\begin{equation}
\begin{aligned}
    \boldsymbol{x}_{t', \mathrm{ave}}^{(i+1)} =& \frac{(\boldsymbol{c}_{\mathrm{att}, 2t}^{(i)} + \boldsymbol{c}_{\mathrm{att}, 2t+1}^{(i)})}{2} \\+&\mathrm{FeedForward}((\boldsymbol{c}_{\mathrm{att}, 2t}^{(i)}, \boldsymbol{c}_{\mathrm{att}, 2t+1}^{(i)})) \label{eq: ave}
\end{aligned}
\end{equation}

\begin{equation}
\begin{aligned}
    \boldsymbol{x}_{t', \mathrm{aff}}^{(i+1)} =& \mathrm{tanh}(\boldsymbol{W}_{\mathrm{aff}}(\boldsymbol{c}_{\mathrm{att},2t}^{(i)}, \boldsymbol{c}_{\mathrm{att}, 2t + 1}^{(i)})+ b_{\mathrm{aff}})\\ +& \mathrm{FeedForward}((\boldsymbol{c}_{\mathrm{att}, 2t}^{(i)}, \boldsymbol{c}_{\mathrm{att}, 2t+1}^{(i)}))\label{eq: aff}
\end{aligned}
\end{equation}

In our experiments, both methods shows close performance. Therefore when showing the results, unless otherwise specified, we use the results of "ave" version.

\ 

\subsection{Decoder and attention mechanisms}

For our GRU models, we use a regular multi-layer uni-directional GRU:
\begin{equation}
    \bar{\boldsymbol{h}}_{t}^{(i)} = \mathrm{GRU}(\bar{\boldsymbol{h}}_{t-1}^{(i)}, \bar{\boldsymbol{h}}_{t}^{(i-1)})
\end{equation}
In our experiment, we did comparison study on Bahdanau Attention(eq. \ref{eq:bahdanau}) and different Luong Attentions. Bahdanau Attention is described in following set of equations.
\begin{align}
    \label{eq:bahdanau}
    u_{tk} & = (\boldsymbol{W}_1\bar{\boldsymbol{h}}_{t}^{(M)} + b_1)^\top(\boldsymbol{W}_2\boldsymbol{h}_{k}^{(N)} + b_2)\\
    \alpha_{tk} & = \frac{u_{tk}}{\sum_ju_{tj}}\\
    \boldsymbol{a}_t & = \sum_j \alpha_{tj}\boldsymbol{h}_{j}^{(N)}\label{eq:context}
\end{align}
$u$ is the alignment score, $h$ and $\bar{\boldsymbol{h}}$ denotes the hidden state in encoder and decoder respectively. $M$, $N$ are the number of layers in decoder and encoder respectively. $a_t$ is the context vector, which will be concatenated with the decoder hidden state of last layer for predicting the next word $\hat{y_t}$.

Luong's global Attentions are generalizations to Bahdanau attention, but using different alignment score calculation methods. For simplicity, we omit the superscript $(M)$ and $(N)$.
\begin{equation}
    u_{tk} = 
    \begin{cases}
        \bar{\boldsymbol{h}}_{t}^\top\boldsymbol{h}_{k} & \mbox{dot}\\
        \bar{\boldsymbol{h}}_{t}^\top\boldsymbol{W_a}\boldsymbol{h}_{k} & \mbox{general}\\
        \boldsymbol{v}_a^\top \mathrm{tanh}(\boldsymbol{W_a} [\bar{\boldsymbol{h}}_{t}, \boldsymbol{h}_{k}]) &\mbox{concat}
    \end{cases}
\end{equation}

We also tried one example of Luong's local attention, which is done by imposing a Gaussian on eq. \ref{eq:context} at a desired attention center $p_t$:
\begin{align}
    \boldsymbol{a}_t & = \sum_j (\alpha_{tj}\boldsymbol{h}_{j})\mathrm{exp}(-\frac{(j-p_t)^2}{2\sigma^2})\\
    p_t & = S\cdot\mathrm{sigmoid}(\boldsymbol{W_p}\bar{\boldsymbol{h}}_t)
\end{align}
where $S$ denotes the total length of the hidden state from the last encoding layer, and $\sigma$ is a parameter chosen manually.

\ 

\subsection{Beam Search}

We use beam search in test and validation where text generation is involved. For each time step, we rank candidate based on their total negative logarithmic probability to current decoding time step $t_{\text{dec}}$:
\begin{equation}
    \text{score} = -\sum_t^{t_\text{dec}} \log(P(\hat{y}))
\end{equation}
The search stop when there are five completed candidates.

\section{Datasets}

We perform our experiments mainly on the Juliet Test Suite for C/C++ (v1.2). (created by \cite{juliet2013c}) This dataset contains 61,387 test cases, each test case contains one flawed code instance and one to several repaired code instance. We note that the instances contains significant amount of dead code. To make the code more realistic, we remove the dead code. We also found that many of the code instances contains if conditions, that in the flawed code instance, it executes one branch, while in the repaired instance, it executes the other. These instances are unrealistic, therefore we removed them. We also performed function re-naming. After the pre-processing, we obtained 31,082 pairs of good-bad code instances. 

To test model's generality, for some of the models, we also tested their performance on Juliet Test Suite for Java (v1.3). (released by \cite{juliet2013j}) After similar pre-processing described above, we obtain 23,015 pairs of instances. 

For both datasets, we divide them up into 80\%, 10\%, 10\% as train, test, and validation dataset respectively.

\section{Results}

\ 

\subsection{Repair Rate}

We train our models on a GeForce GTX 1080 Ti graphic card. The metric we use for evaluation is the repair rate, which is the fraction of instances that are repaired after the model's edit.  Since we performed beam search with beam width 5, each time a correction is being performed, we generate 5 correction candidates. Here we have two metrics in measuring the performance: 1 candidate repair rate and 5 candidate repair rate. The former corresponds to the scenario of code auto-correction, where there is no human judgement involved. The latter corresponds to correction suggesting, where the machine will identify an error and provide suggestions for the programmer for futher judgement. The comparisons of the repair rates for considered models and their counterparts with pyramid encoder are listed in Table~\ref{tab:repair_C} and Table~\ref{tab:repair_Java}.

From these results we see, Pyramid encoder have close performance to regular encoder in most of the models we applied to, except for with Luong's local attention. The reason is that the encoder output in pyramid encoder is very "coarse-grained", each output position now represents information from $2^{(N-1)}$ words. This results in two drawbacks specifically to local attention: one, a much more "blurry" attention center; two, a much broader attention window. As a result, the attention is much less targeted, which damages the performance. Therefore, in the rest of the paper, we will exclude this attention mechanism from our discussion.


\begin{table}[]
\caption{Repair rate of GRU and Transformer on Juliet Test Suite for C/C++, comparing the regular encoder and pyramid encoder. Apparently pyramid encoder does not collaborate well with Luong's local attention, therefore we exclude it from future discussions. It is also not included when calculating the average improvement.}
\centering
\footnotesize
\vskip 0.1in
\label{tab:repair_C}
\resizebox{\textwidth}{!}{%
\begin{tabular}{|l|c|c|c|c|}
\hline
\multicolumn{1}{|c|}{\multirow{2}{*}{model}} & \multicolumn{2}{c|}{1 candidate repair rate (\%)}                           & \multicolumn{2}{c|}{5 candidate repair rate(\%)}          \\ \cline{2-5} 
\multicolumn{1}{|c|}{}                       & \multicolumn{1}{l|}{Regular encoder} & \multicolumn{1}{l|}{Pyramid encoder} & Regular encoder & Pyramid encoder                         \\ \hline
GRU+Bahdanau Att                             & \textbf{75.76}      & 73.24 (-2.52)                        & 94.19           & 95.99 (+1.80)                           \\
GRU+Luong Att: dot                           & 73.18                                & 69.36 (-3.82)                        & 93.94           & 94.10 (+0.16)                           \\
GRU+Luong Att: general                       & 72.57                                & 72.99 (+0.42)                        & 94.00           & 94.55 (+0.55)                           \\
GRU+Luong Att: concat                        & 50.34                                & 47.26 (-3.08)                        & 86.72           & 86.14 (-0.58)                           \\
GRU+Luong Att: local                & 65.70           & 49.18 (-15.52)             & 92.46                    & 86.24 (-6.22)    \\
Transformer                                  & 74.26                                & 72.39 (-1.87)                        & 95.60           & \textbf{96.37} (+0.77) \\
\hline
Average Improvement (\%) & \multicolumn{2}{c|}{-2.17} & \multicolumn{2}{c|}{+0.54}\\
\hline
\end{tabular}%
}
\end{table}

\begin{table}[]
\caption{Repair rate of GRU and Transformer on Juliet Test Suite for Java, comparing the regular encoder and pyramid encoder.}
\centering
\footnotesize
\vskip 0.1in
\label{tab:repair_Java}
\resizebox{\textwidth}{!}{%
\begin{tabular}{|l|c|c|c|c|}
\hline
\multicolumn{1}{|c|}{\multirow{2}{*}{model}} & \multicolumn{2}{l|}{1 candidate repair rate (\%)}                           & \multicolumn{2}{c|}{5 candidate repair rate(\%)} \\ \cline{2-5} 
\multicolumn{1}{|c|}{}                       & \multicolumn{1}{l|}{Regular encoder} & \multicolumn{1}{l|}{Pyramid encoder} & Regular encoder     & Pyramid encoder            \\ \hline
GRU+Bahdanau Att                             & 54.88                                & 56.34 (+1.46)                        & 84.22               & 84.11 (-0.11)              \\
GRU+Luong Att: dot                           & 55.10                                & 55.59 (+0.49)                        & 82.54               & 84.56 (+2.02)               \\
GRU+Luong Att: general                       & 52.76                                & 52.54 (-0.22)                        & 82.67               & 82.62 (-0.05)              \\
GRU+Luong Att: concat                        &                                      &                                      &                     &                            \\
Transformer                                  & 56.22                                & \textbf{56.76 (+0.54)}               & 92.97               & \textbf{93.68 (+0.71)}  \\
\hline
Average Improvement (\%) & \multicolumn{2}{c|}{+0.58} & \multicolumn{2}{c|}{+0.64}\\
\hline
\end{tabular}%
}
\end{table}

\ 

\subsection{Converging Speed}

Since Pyramid encoder reduces the sequence lengths in higher layers, one can expect a smaller training cost per batch in both GRU and Transformer models. To quantify this effect, for each of the regular encoder-pyramid encoder model pairs in Table\ref{tab:repair_C}, we set the batch size the same, and compare the average training speed in words per second, as shown in Table \ref{tab:speed}. Here the batch size is chosen so that it optimizes the training speed on the given GPU for each model. In the model, we also included the number of epochs for the model to converge.

\begin{table}[]
\caption{Training speed of GRU and Transformer on Juliet Test Suite for C/C++}
\centering
\footnotesize
\vskip 0.1in
\label{tab:speed}
\resizebox{\textwidth}{!}{%
\begin{tabular}{|l|c|c|c|c|c|}
\hline
\multicolumn{1}{|c|}{\multirow{2}{*}{model}} & \multirow{2}{*}{Batch Size} & \multicolumn{2}{c|}{\begin{tabular}[c]{@{}c@{}}Training Speed \\ (words/s)\end{tabular}} & \multicolumn{2}{c|}{Converge Epoch} \\ \cline{3-6} 
\multicolumn{1}{|c|}{}                       &                             & Regular                                 & Pyramid                                        & Regular          & Pyramid          \\ \hline
GRU+Bahdanau Att                             & 8                           & 754                                     & 1185 (+57\%)                                   & 18               & 18               \\
GRU+Luong Att: general                       & 16                          & 441                                     & 853 (+108\%)                                   & 23               & 27               \\
GRU+Luong Att: dot                           & 128                         & 4646                                    & 10408 (+124\%)                                 & 36               & 34               \\
GRU+Luong Att: concat                        & 6                           & 1418                                    & 2344 (+65\%)                                   & 23               & 29               \\
Transformer                                  & 8                           & 1086                                    & 2181 (+101\%)                                   & 33               & 34               \\ \hline
\end{tabular}%
}
\end{table}

\begin{figure}
    \centering
    \includegraphics[width=0.6\textwidth]{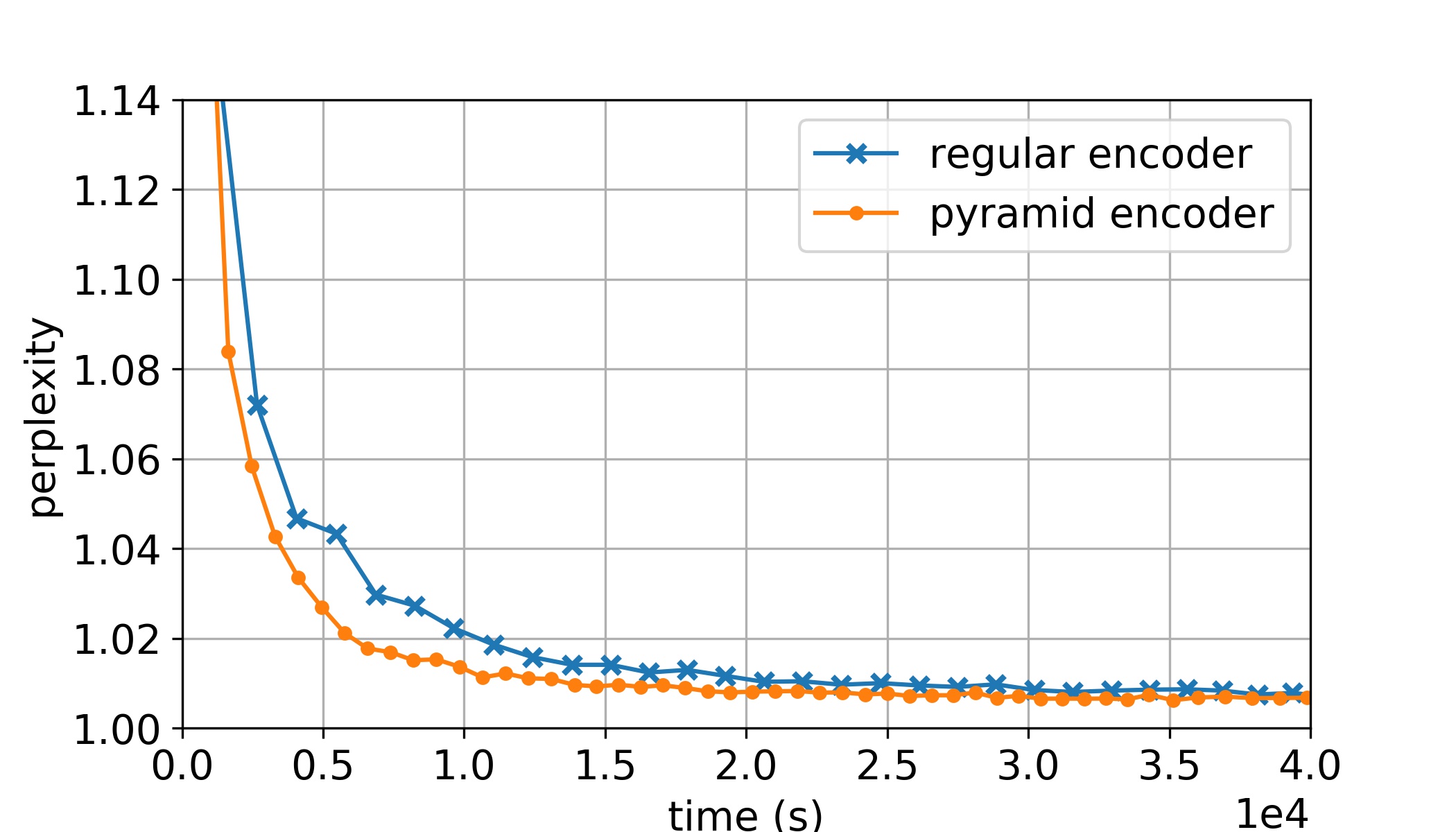}
    \caption{Learning curve of GRU with Luong's general attention, comparing regular encoder to pyramid encoder. Pyramid encoder model shows fast converging speed.}
    \label{fig:learning_curve}
\end{figure}

Apparently it takes similar number of epochs to converge for the same type of model with pyramid encoder and regular encoder. However, pyramid encoders largely increases the training speed, between 50\% to 130\%. Therefore it could easily shorten the training time by two to four folds while same performance is achieved. As an example, Figure~\ref{fig:learning_curve} shows the learning curve for GRU model with general Luong's attention, comparing the regular encoder and pyramid encoder.

\ 

\subsection{Memory cost}

The last thing we compared is the memory cost of the pyramid encoder and the regular encoder. This measure is crucial in some scenario, where your input instances are very long, therefore the memory of GPU is only capable of holding a very small batch. In code correction, this is often the case.

The metric we use for comparison is memory cost per instance, $k$, which is defined as 
\begin{equation}
    k = \frac{\Delta \text{Memory usage}}{\Delta \text{Batch size}}
\end{equation}

\begin{figure}
    \centering
    \includegraphics[width=0.5\textwidth]{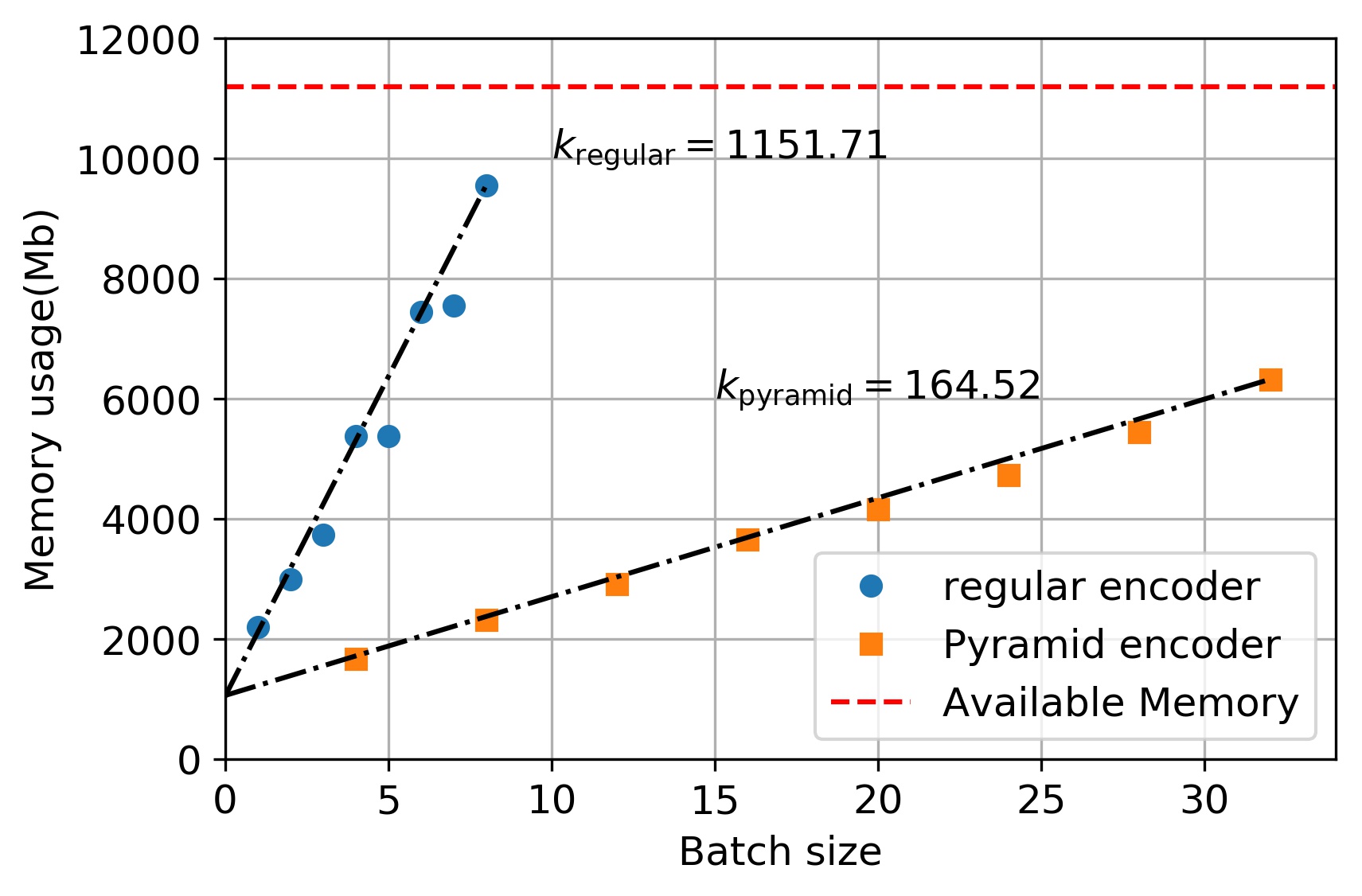}
    \caption{Memory cost per instance for GRU models with Bahdanau attention, $k$ is calculated by finding the slope of the linear fit (black dashed line). The red dashed line represents the maximum memory of a GeForce GTX 1080 Ti graphic card.}
    \label{fig:MemoryUse}
\end{figure}

Figure~\ref{fig:MemoryUse} shows the calculation process of $k$. Define $\mathcal{E} = 1/k$ as memory efficiency, We calculated the $k$ and $\mathcal{E}$ value for each of the models we applied, shown in Table~\ref{tab:memory}. We also included number of parameters in each model. From which we see that each pair of models have roughly the same model size.

The pyramid encoder could increase the memory efficiency by 20\% to 600\% depending on the attention mechanisms used, while only increase the memory occupied by the model itself by around 10\%. One should note that the memory efficiency directly effects the maximum batch size one is able to use on a single GPU, and therefore effects the utility of the GPU. For example, for regular GRU with Bahdanau Attention, given an input length of instances being 500, a GeForce GTX 1080 Ti graphic card could only carry out training with batch size of 8, this and does not fully utilize the GPU. with Pyramid encoder, it's capacity goes up to 60 instances each batch. In practice, this will drastically reduce the training time by increasing the GPU utility, together with the smaller computational cost of pyramid encoder as addressed in previous section.

\begin{table}[]
\caption{Memory Cost for considered models, comparing regular encoder and pyramid encoder, pyramid encoder greatly increased the memory efficiency.} .

\centering
\footnotesize
\vskip 0.1in
\label{tab:memory}
\resizebox{\textwidth}{!}{%
\begin{tabular}{|l|c|c|c|c|c|c|}
\hline
\multicolumn{1}{|c|}{\multirow{2}{*}{model}} & \multicolumn{2}{c|}{k(Mb/instance)} & \multicolumn{2}{c|}{$\mathcal{E}$($10^{-3}$)} & \multicolumn{2}{c|}{parameters ($10^7$)} \\ \cline{2-7} 
\multicolumn{1}{|c|}{}                       & Regular    & Pyramid   & Regular    & Pyramid         & Regular           & Pyramid           \\ \hline
GRU+Bahdanau Att                             & 1151.71    & 164.52    & 0.86       & 6.08 (+600\%)   & 1.24              & 1.11              \\
GRU+Luong Att: general                       & 830.71     & 165.03    & 1.20       & 6.05 (+403\%)   & 1.22              & 1.10              \\
GRU+Luong Att: dot                           & 65.91      & 52.42     & 15.17      & 19.08 (+26\%)   & 1.20              & 1.08              \\
GRU+Luong Att: concat                        & 1381.6     & 431.87    & 0.72       & 2.31 (+220\%)   & 1.24              & 1.11              \\
Transformer                                  & 414.67     & 263.33    & 0.24       & 0.38 (+57\%)    & 2.35              & 2.82              \\ \hline
\end{tabular}%
}
\end{table}

\section{Discussion}

\ 

\subsection{Length analyses}

\begin{figure}
    \centering
    \includegraphics[width=0.6\textwidth]{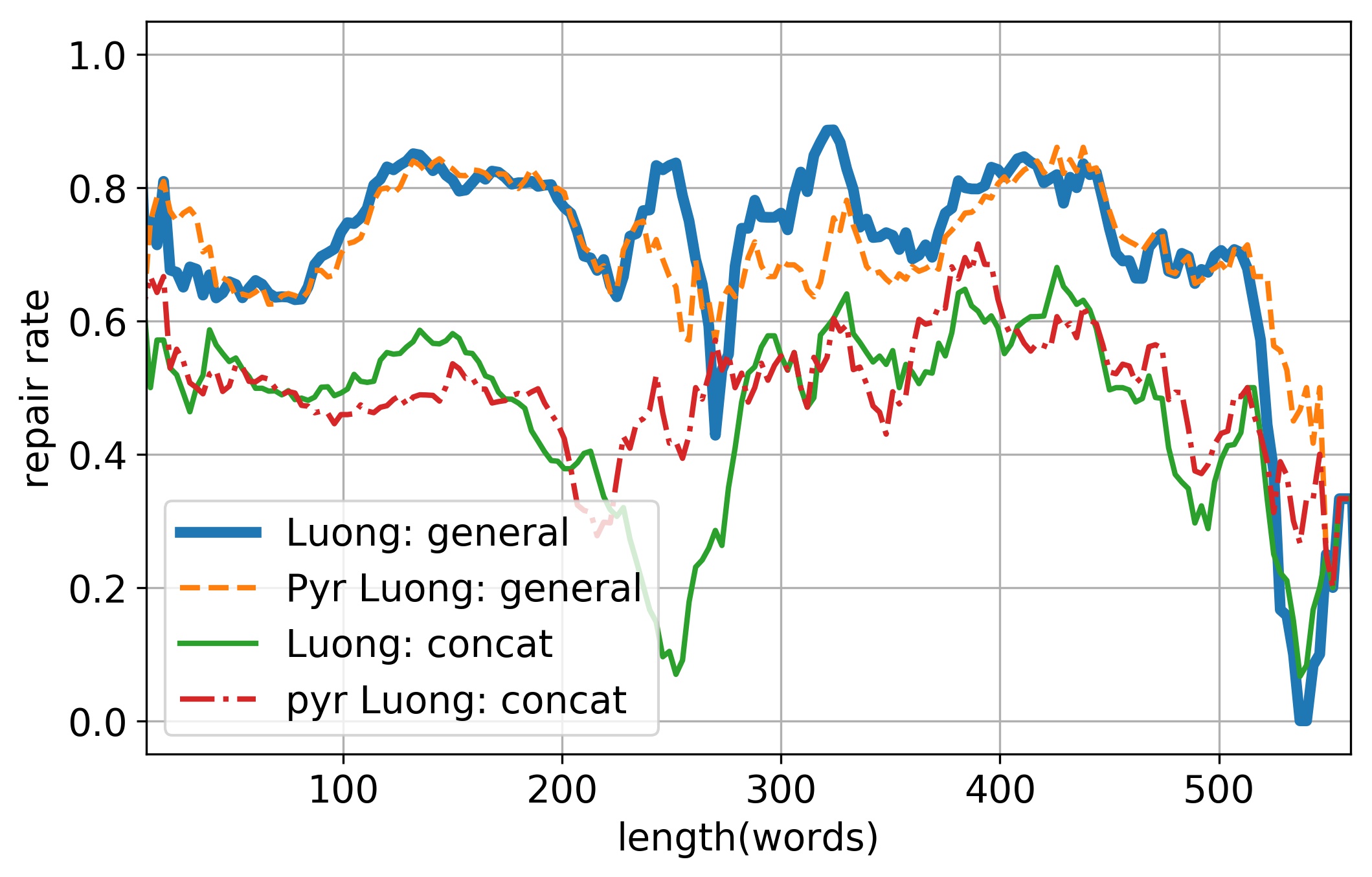}
    \caption{Length analyses of Luong's general attention and Luong's concat attention. The results from the rest of the models are qualitatively similar to result of Luong's general attention, thus omitted. }
    \label{fig:LengthAnalyses}
\end{figure}

\begin{figure}
    \centering
    \includegraphics[width=0.95\textwidth]{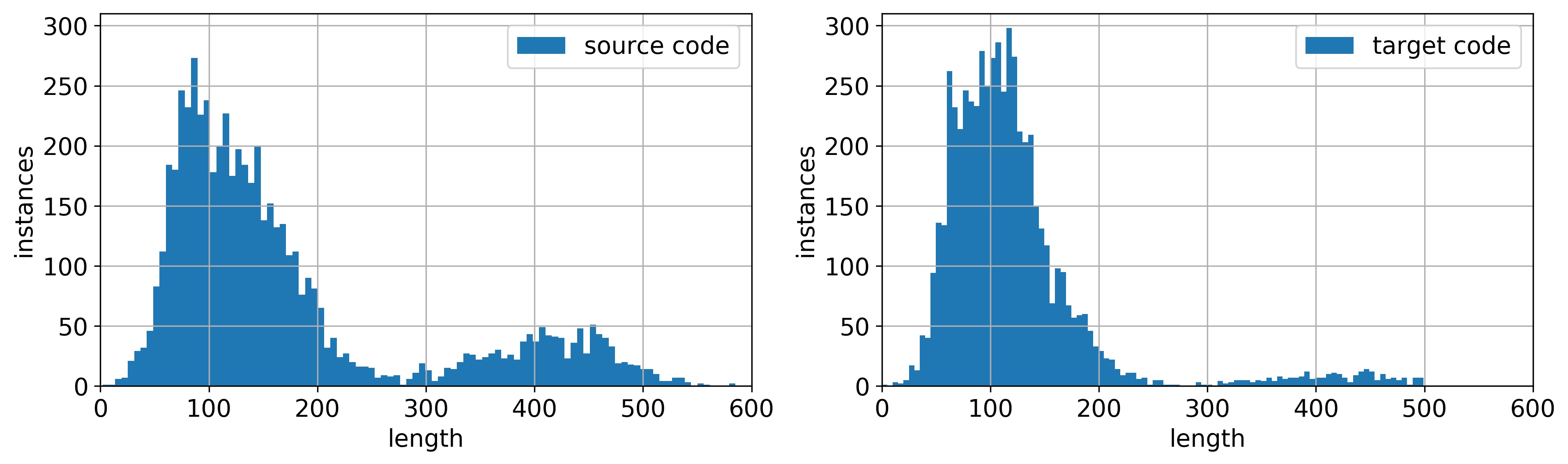}
    \caption{Histogram of flawed code (left) and repaired code (right) instances. }
    \label{fig:histogram}
\end{figure}

Figure \ref{fig:LengthAnalyses} shows the repair rate of the models with respect to the input length. We omitted the result of Transformer, Bahdanau's attention and Luong's general attention, because they are qualitatively similar to the result of Luong's dot attention. Despite the different attention mechanisms, these seq2seq models (with pyramid encoder or regular encoder) are relatively robust to longer input lengths. The performance drop at around 250 words and above 500 words are likely resulting from the shortage of samples, which one can easily observe from Figure \ref{fig:histogram}, the length histogram of source instances and target instances. The histogram also shows that the majority of code instances contains several hundred words, while natural language sentences are typically not longer than 50 words. This feature of code instances calls for a much higher computational resource requirement for PLC problems than NLC problems, which makes pyramid structure especially useful.

\ 

\subsection{Examples of correction}

In this section we give several examples of successful corrections from our Pyramid GRU model on Juliet C/C++ Test Suite for closer examination of model and datasets. The red striked out texts denote the original faulted instance, blue buffed texts are the reparation done by the model. 

Example 1: Memory allocation match

The flawed code creates a char variable whose size does not match its concatenating destination. The model is able to correct it so that their size matches each other.
\begin{MyVerbatim}[commandchars=\\\{\}]
void main()\{
    char * data;
    char data_buf[100];
    data=data_buf;
    \textcolor{red}{\sout{memset(data,'A', 100-1);}}
    \textcolor{blue}{\textbf{memset(data,'A', 50-1);}}
    data[100-1]='\textbackslash0'; 
    \{
        char dest[50]="";
        strncat(dest,data,strlen(data));
        dest[50-1]='\textbackslash0'; 
        printLine(data);
    \}
\}
\end{MyVerbatim}

Example 2: Redundant code

This is an example that the model deletes repeated code where a variable is freed twice.

\begin{MyVerbatim}[commandchars=\\\{\}]
void main()\{
    char * data;
    data=NULL;
    data=(char *)malloc(100*sizeof(char));
    free(data);
    \textcolor{red}{\sout{free(data);}}
\}
\end{MyVerbatim}

Example 3: Possible Overflow

Here we show a slightly questionable example of correction provided by the dataset. In order to prevent potential string overflow emerging from environment variable, the repair suggestion given by the Juliet Test Suite is to abort the entire part of concatenating the environment string and replace the variable with an arbitrary string "*.*". This "correction" is easy for the model to learn, however, it has changed the original purpose of the program. 

\begin{MyVerbatim}[commandchars=\\\{\}]
void main()\{
    char * data;
    char data_buf[100]="";
    data=data_buf;
    \textcolor{red}{\sout{size_t data_len=strlen(data);}}
    \textcolor{red}{\sout{char * environment=GETENV(ENV_VARIABLE);}}
    \textcolor{red}{\sout{if(environment!=NULL)\{}}
        \textcolor{red}{\sout{strncat(data+data_len,}}
                \textcolor{red}{\sout{environment,100-data_len-1);}}
    \textcolor{red}{\sout{\}}}
    \textcolor{blue}{\textbf{strcat(data,"*.*");}}
    _spawnl(_P_WAIT,COMMAND_INT_PATH,
            COMMAND_INT_PATH,
            COMMAND_ARG1,COMMAND_ARG2,
            COMMAND_ARG3,NULL);
\}
\end{MyVerbatim}

Example 4: Correction across functions

In this example, the models demonstrates the ability of making connections across the whole instance, between different functions. Here it prevents potential overflow in the sink function caused by a variable that was passed from the main function by adding an if condition.

\begin{MyVerbatim}[commandchars=\\\{\}]
static void sink(unsigned char data)
\{
    \textcolor{blue}{\textbf{if (data < UCHAR_MAX)\{}}
        unsigned char result = data + 1;
    \textcolor{blue}{\textbf{\}}}
\}
void main()
\{
    unsigned char data;
    data = ' ';
    data = (unsigned char)rand();
    sink(data);
\}

\end{MyVerbatim}

\ 

\subsection{Alternative method for small datasets: Transfer Learning}

One main difficulty that researchers often come across when attempting to apply machine learning methods to PLC problems is the availability of suitable datasets. Although there are many datasets and shared tasks available on \cite{datasets}, most of them includes less than 1,000 examples. This makes neural network based methods nearly impossible. To tackle this problem, we take the idea of transfer learning from \cite{pan2009survey}. 

Our idea is to take the encoder part of the model that was trained on Juliet Test Suite, and attach it to a untrained decoder, which was designed for the specific problem. We aim to take the advantage that codes written in the same coding language shares the same syntax library and same construction rules. 

Since many datasets available only provide the faulted code and their corresponding fault categories, here we give an example of fault classification using transfer learning, applying the model pre-trained on Juliet Test Suite for C/C++ on ITC benchmark (\cite{itc2015}).

\ 

\subsubsection{Model structure}

Given a faulted code instance, our goal is to train a classification model that predicts the type of error of the faulted code from a given list of error categories. 

We keep the encoder part of the pre-trained model and use it directly as the encoder in the classification problem. The exception is the embedding layer, because the vocabulary in the new dataset will contain new variable names that did not occur in pre-trained embedding, although the syntax will be the same. In practice, we manually expanded the embedding layer to accommodate the new "words", but keep the embedding of the old "words" unchanged. In order to add variation from the original model, we also re-initialized the weights in the last encoding layer. 

For the decoder, instead of generating a sequence, we take the output of the first time step of the re-initiated decoder, pass it to a linear layer that project it to an $n_{\text{class}}$ dimensional vector. $n_{\text{class}}$ is the number of error classes. Model was trained on minimizing cross-entropy loss with ADAM optimizer.

\ 

\subsubsection{Results}

We extracted 566 C/C++ code instances from the ITC bench mark. These instances are organized into 44 error categories, with the largest category containing around 30 instance and the smallest only containing 2 instances. Then the instances are divided into a training set of 485 instances, a validation set of 42 instances and test set of 39 instances. For comparison, we also tried Pyramid GRU and Pyramid Transformer with same model structure but no prior knowledge from Juliet Test Suites. The result is shown in Table \ref{tab: TFL}.

For the fresh GRU and Transformer models, we observed that the models has no predicting power as it produces constant prediction over all inputs. There is even no sufficient gradient on the loss landscape as the loss did not reduce during the training. Transfer learning, on the other hand, demonstrates a fair power of prediction, correctly classifies over 60\% of instances, despite that ITC benchmark is written in very different style than Juliet Test Suites and that the dataset is 50 times smaller.

This result shows that one is able to use neural-network-based methods in code correction problems despite the shortage of data, which is a common problem in this field.

\begin{table}[]
\caption{compare result of transfer learning on error type classification task. The models without transfer learning demonstrate no predicting power and no improvement during course of training.}

\centering
\footnotesize
\vskip 0.1in

\label{tab: TFL}
\resizebox{0.45\textwidth}{!}{%
\begin{tabular}{l|l}
model                     & accuracy (\%) \\ \hline
Transfer Learning: PyrGRU & 60.5          \\
Transfer Learning: PyrTFM & \textbf{69.1} \\
Fresh pyramid GRU               & 16.7          \\
Fresh pyramid Transformer         & 7.1          
\end{tabular}%
}
\end{table}

\section{Conclusion}

In our work, we show that seq2seq models, successful in natural language correction, is also applicable in programming language correction. Our result shows seq2seq models can be well applied in providing suggestions to potential errors, and have a decent correct rate (above 70\% in C/C++ dataset and above 50\% in Java dataset) in code auto-correction. Although these results are only limited in Juliet Test Suites, we expect that given enough training data, seq2seq models can also perform well when applied on other PLC problems. 

Based on the common used encoder-decoder structure, We introduce a general Pyramid Encoder in seq2seq models. Our result demonstrates that this structure significantly reduces the memory cost and computational cost. This is helpful because PLC are generally more computationally expensive than NLC, due to its longer average instance length.

The publicly available datasets in PLC are mostly small and noisy. Most datasets we found contains close to or less than 1000 code instances. This is far less than enough for training seq2seq and many other machine learning models. Our results on transfer learning pointed out a way of processing these small dataset using pre-trained model as an encoder, which boosts the performance by a large amount.

In future, we will further investigate the influence of different architectures in neural networks; for instance, parallel encoder/decoders, Tree2Tree models, etc. On the other hand, instead of code correction, we will modify and examine our model's performance on other tasks such as program generation, code optimising, etc. We will also examine the potential difference between artificial datasets and realistic datasets.

\bibliography{CodeArxiv}

\bibliographystyle{unsrt}  


\end{document}